\documentclass[12pt,a4paper]{article}
\pdfoutput=1
\usepackage[utf8]{inputenc}
\usepackage{epsf,amsmath,amssymb,graphicx,dcolumn}
\usepackage{caption}
\usepackage{subcaption}
\usepackage{scalefnt,ulem,pstricks}
\usepackage{booktabs,multirow,tabularx}
\usepackage{cite,hyperref}
\usepackage{cleveref}
\usepackage{color}
\usepackage{rotating}
\usepackage{microtype}
\usepackage[titletoc,title]{appendix}
\usepackage[numbers,sort&compress]{natbib}
\usepackage{amsmath,amsfonts,amsthm,bm} % Math packages
\usepackage[tikz]{bclogo}
\usepackage{subcaption}
\usepackage{tikz-feynman}
 
\newcommand\F{${\rm F}$}
\newcommand\FJ{${\rm FJ}$}
\newcommand\FJJ{${\rm FJJ}$}
\newcommand\PhiB{\Phi_{\scriptscriptstyle \rm F}}

\newcommand\PhiBJ{\Phi_{\scriptscriptstyle \rm FJ}}

\newcommand{\pt}{{p_{\text{\scalefont{0.77}T}}}}
\newcommand{\ptrad}{{p_{\text{\scalefont{0.77}T,rad}}}}

\newcommand{\muF}{{\mu_{\text{\scalefont{0.77}F}}}}
\newcommand{\muR}{{\mu_{\text{\scalefont{0.77}R}}}}

\newcommand{\noun}[1]{{\tt #1}}

\newcommand{\POWHEG}{\noun{POWHEG}}

\newcommand{\minlo}{{\noun{MiNLO$^{\prime}$}}}
\newcommand{\minnlo}{{\noun{MiNNLO$_{\rm PS}$}}}
\newcommand{\Matrix}{{\noun{Matrix}}}
\newcommand{\PYTHIA}[1]{\noun{Pythia{#1}}}

\newcommand{\radish}{\noun{RadISH}}

\newcommand{\abarmu}[1]{\frac{\as(#1)}{2\pi}}

\newcommand{\nnlops}{NNLO+PS}

\newcommand{\citere}[1]{Ref.\,\cite{#1}}

\newcommand{\fig}[1]{Fig.\,\ref{#1}}

\newcommand{\LambdaPWG}{\Lambda_{\rm pwg}}

\newcommand{\ptvec}{{\vec{p}_T}}
\DeclareMathOperator{\Tr}{Tr}

\usepackage{xcolor}
\newcommand{\mathd}{\mathrm{d}}
\newcommand{\tmop}[1]{\ensuremath{\operatorname{#1}}}

\newcommand{\sphid}[1]{}

\providecommand{\href}[2]{#2}

\newcommand\as{\alpha_{\mathrm{S}}}

\def\to{\rightarrow}

\def\citere#1{\mbox{Ref.~\cite{#1}}}

\makeatletter
\g@addto@macro\bfseries{\boldmath}
\makeatother

\begin{document} 
\begin{flushright}
\vspace*{-1.5cm}
%PREPRINT \\
\end{flushright}
\vspace{0.cm}

\begin{center}
  {\Large \bf NNLO+PS with M{\scalefont{0.76}I}NNLO$_{\text{PS}}$: status and prospects}\\[0.5cm]

\end{center}

\begin{center}
  {\bf Luca Buonocore}$^{(a)}$,
  {\bf Mauro Chiesa}$^{(b)}$,  
  {\bf Gabri\"el Koole}$^{(c)}$,
  {\bf Daniele Lombardi}$^{(c)}$,
  {\bf Javier Mazzitelli}$^{(c)}$,  
  {\bf Pier Francesco Monni}$^{(d)}$,
  {\bf Paolo Nason}$^{(c,e)}$,
  {\bf Emanuele Re}$^{(e,f)}$,
  {\bf Luca Rottoli}$^{(a)}$,
  {\bf Marius Wiesemann}$^{(c)}$
  {\bf Giulia Zanderighi}$^{(c)}$,
  {\bf Silvia Zanoli}$^{(c)}$

$^{(a)}$ University of Z\"urich, Winterthurerstrasse 190, 8057 Z\"urich, Switzerland\\
$^{(b)}$ Universit\`a di Pavia and INFN, Sezione di Pavia, Via A. Bassi 6, 27100 Pavia, Italy\\  
$^{(c)}$ Max-Planck-Institut f\"ur Physik, F\"ohringer Ring 6, 80805
M\"unchen, Germany\\
$^{(d)}$ CERN, Theoretical Physics Department, CH-1211 Geneva 23, Switzerland\\
$^{(e)}$ Universit\`a di Milano\,-\,Bicocca and INFN, Sezione di Milano\,-\,Bicocca, Piazza della Scienza 3, 20126 Milano, Italy\\
$^{(f)}$ LAPTh, Universit\'e Grenoble Alpes, USMB, CNRS, 74940 Annecy, France
\end{center}

\begin{abstract}

We summarize the current status and near future prospects for
next-to-next-to-leading order calculations matched to parton shower
based on the \minnlo{} method. We give a theoretical overview, illustrate
selected results for $ZZ\to 4\ell$ and top-pair production processes
at the LHC, and provide an outlook of the future challenges.

\end{abstract}

\section{Introduction}

The developments in the theoretical description of collider physics
processes have seen a steady improvement in the precision of
perturbative calculations, also matched to parton shower
simulations. While twenty years ago the standard of Monte Carlo event
simulations was just leading order (LO), today it is instead
next-to-leading-order (NLO). However, the requirement to match the
precision of LHC measurements, often reaching percent-level precision,
has fostered developments towards the next-to-next-to-leading order
(NNLO) level. Ten years have passed since the very first NNLO+parton
shower (NNLOPS) implementations for Higgs and Drell-Yan production
appeared, and new methods have been proposed which allowed a leap in
the complexity of the processes that can be described. It is not
unlikely that by the time the high-luminosity program will start,
NNLOPS will constitute the new precision standard, together with an
improved precision of the logarithmic accuracy in the parton shower to
full next-to-leading-logarithmic (NLL) or even
next-to-next-to-leading-logarithmic (NNLL).

The consistent combination of next-to-next-to-leading order (NNLO) QCD
calculations with parton-shower simulations (\nnlops{}) is one of the
current challenges in collider theory, and it is indispensable to
provide the interface between accurate theory predictions and
precision measurements.
A good \nnlops{} method should attain NNLO
accuracy for observables inclusive in the QCD radiation beyond the
Born level, while preserving the logarithmic structure (and accuracy)
of the parton-shower simulation after matching.

In \citere{Monni:2019whf,Monni:2020nks} we have presented the method
\minnlo{}, which is an extension of the \minlo{} procedure of
Refs.~\cite{Hamilton:2012np,Hamilton:2012rf}. The latter method was
used to obtain NNLOPS accuracy using a reweighting procedure.
It was applied to some simple LHC processes, namely
Higgs-boson production~\cite{Hamilton:2013fea}, the
Drell-Yan
process~\cite{Karlberg:2014qua} and Higgs to bottom quark decays~\cite{Bizon:2019tfo}, and more
complicated LHC processes, such as the two Higgs-strahlung
reactions~\cite{Astill:2016hpa,Astill:2018ivh}, and the production of
two opposite-charge leptons and two neutrinos
($W^+W^-$)~\cite{Re:2018vac}. 
These computations have employed the reweighting procedure to its
extreme.
In contrast, the new \minnlo{} procedure
of~\citere{Monni:2019whf,Monni:2020nks} addresses more directly the
requirement of NNLO accuracy and thus, besides not using any
reweighting, it can be more easily generalized to processes beyond
massive colour-singlet production.  It meets the following features:
\begin{itemize}
\item NNLO corrections are calculated directly during the generation
  of the events and without any additional a-posteriori reweighting. 
\item No merging scale is required to separate different multiplicities in the
  generated event samples.
\item The matching to the parton shower is performed according to the
  \POWHEG{} method~\cite{Nason:2004rx,Frixione:2007vw,Alioli:2010xd} and preserves the leading 
  logarithmic (LL) structure of transverse-momentum ordered 
  showers.\footnote{For a different ordering variable, preserving the
    accuracy of the shower is more subtle. Not only one needs to veto
    shower radiation that has relative transverse momentum greater
    than the one generated by \POWHEG{}, but also one has to resort to
    truncated showers~\cite{Nason:2004rx,Bahr:2008pv} to compensate
    for missing collinear and soft radiation. Failing to do so spoils the
    shower accuracy at leading-logarithmic level (in fact, at the
    double-logarithmic level).}
\end{itemize}
This method, that up to now has been applied to the production of
colour singlet system such as
$Z\gamma$~\cite{Lombardi:2020wju,Lombardi:2021wug},
$W^+W^-$~\cite{Lombardi:2021rvg}, $ZZ$~\cite{Buonocore:2021fnj}, $VH$
including the $H\to b\bar b$ decay at NNLOPS~\cite{Zanoli:2021iyp},
was extended to deal with the production of massive coloured final
states in Refs.~\cite{Mazzitelli:2020jio,Mazzitelli:2021mmm}, where it
was applied to top-quark pair production. This was the first NNLOPS result for
LHC processes with colored final states.

In this manuscript we first give an overview of the theoretical
aspects at the core of \minnlo{} NNLO+PS simulations.
%(section~\ref{sec:th}).
We then discuss selected results for $ZZ$ and $t\bar t$ production, and conclude with an outlook of possible future developments.
%in section~\ref{sec:pheno}.

\section{Theoretical overview}
\label{sec:th}

\subsection{M{\scalefont{0.76}I}NNLO$_{\text{PS}}$ in a nutshell}
\label{sec:minnlo}
The \minnlo~method~\cite{Monni:2019whf,Monni:2020nks} formulates a
NNLO calculation fully differential in the phase space $\PhiB$ of the
produced colour singlet \F{} with invariant mass $Q$. It starts from a
differential description of the production of the colour singlet and a
jet (\FJ{}), whose phase space we denote by $\PhiBJ$.
It is described by the following formula:
\begin{align}
\frac{\mathd\sigma}{\mathd\PhiBJ}={\bar B}(\PhiBJ) \times
\bigg\{\Delta_{\rm pwg} (\LambdaPWG) + \int\mathd \Phi_{\tmop{rad}} 
  \Delta_{\rm pwg} (\ptrad)  \frac{R (\PhiBJ{}, \Phi_{\tmop{rad}})}{B
  (\PhiBJ{})}\bigg\}\,,
\label{eq:master}
\end{align}
which corresponds precisely to a standard \POWHEG{}
calculation~\cite{Nason:2004rx,Frixione:2007vw,Alioli:2010xd} for
\FJ{} production, supplemented with a modified ${\bar B}(\PhiBJ)$
function that is crucial to reach NNLO accuracy in the production of
the system \F{}.
The function ${\bar B}(\PhiBJ)$ describes the generation of the first
radiation, while the content of the curly brackets describes the
generation of the second radiation according to the
\POWHEG{}~method~\cite{Nason:2004rx,Frixione:2007vw,Alioli:2010xd}.
Here, $B$ and $R$ are the squared tree-level matrix elements for \FJ{}
and \FJJ{} production, respectively.  $\Delta_{\rm pwg}$ denotes the
\POWHEG{} Sudakov form factor~\cite{Nason:2004rx} and
$ \Phi_{\tmop{rad}} $ ($\ptrad$) is the phase space (transverse
momentum) of the second radiation, which is generated above the
\POWHEG{} infrared cutoff $\LambdaPWG = 0.89$~GeV.
The parton shower then adds additional radiation to the partonic
events generated according to Eq.~\eqref{eq:master}, and it
contributes beyond $\mathcal{O}(\as^2(Q))$ at all orders in
perturbation theory.

The function ${\bar B}(\PhiBJ)$ is the central ingredient of
\minnlo{}. Its derivation~\cite{Monni:2019whf} stems from the
observation that the NNLO cross section differential in the transverse
momentum of the color singlet ($\pt$) and in the Born phase space
$\PhiB$ is described by the following formula
\begin{align}
\label{eq:start}
  \frac{\mathd\sigma}{\mathd\PhiB\mathd \pt} &= \frac{\mathd}{\mathd \pt}
     \bigg\{ \exp[-\tilde{S}(\pt)] {\cal L}(\pt)\Bigg\} +
                                               R_f(\pt) =
  \exp[-\tilde{S}(\pt)]\left\{
                                  D(\pt)+\frac{R_f(\pt)}{\exp[-\tilde{S}(\pt)]}\right\}\,,
\end{align}
where $R_f$ contains terms that are integrable in the $\pt\rightarrow 0$ limit, and 
\begin{equation}
\label{eq:Dterms}
  D(\pt)  \equiv -\frac{\mathd \tilde{S}(\pt)}{\mathd \pt} {\cal L}(\pt)+\frac{\mathd {\cal L}(\pt)}{\mathd \pt}\,.
\end{equation}
$\tilde{S}(\pt)$ represents the Sudakov form factor, while ${\cal L}(\pt)$
contains the parton luminosities, the squared virtual matrix elements
for the underlying \F{} production process up to two loops as well as
the NNLO collinear coefficient functions. Explicit expressions can be found in \citere {Monni:2019whf,Monni:2020nks}.
A crucial feature of the \minnlo{} method is that the renormalisation
and factorisation scales are set to $\muR\sim\muF\sim \pt$.

Introducing the NLO differential cross section for \FJ{} production
\begin{equation}
\label{eq:NLO}
\frac{\mathd\sigma^{\rm (NLO)}_{\scriptscriptstyle\rm FJ}}{\mathd\PhiB\mathd
      \pt} = \abarmu{\pt}\left[\frac{\mathd\sigma_{\scriptscriptstyle\rm FJ}}{\mathd\PhiB\mathd
      \pt}\right]^{(1)} + \left(\abarmu{\pt}\right)^2\left[\frac{\mathd\sigma_{\scriptscriptstyle\rm FJ}}{\mathd\PhiB\mathd
      \pt}\right]^{(2)}\,,
\end{equation}
where $[X]^{(i)}$ denotes the coefficient of the $i$-th term in the
perturbative expansion of the quantity $X$, one can rewrite
Eq.~\eqref{eq:start} as
\begin{align}
\label{eq:minnlo}
  \frac{\mathd\sigma}{\mathd\PhiB\mathd \pt}  &=
  \exp[-\tilde{S}(\pt)]\bigg\{ \abarmu{\pt}\left[\frac{\mathd\sigma_{\scriptscriptstyle\rm FJ}}{\mathd\PhiB\mathd
      \pt}\right]^{(1)} \left(1+\abarmu{\pt} [\tilde{S}(\pt)]^{(1)}\right)
  \notag
+ \left(\abarmu{\pt}\right)^2\left[\frac{\mathd\sigma_{\scriptscriptstyle\rm FJ}}{\mathd\PhiB\mathd
      \pt}\right]^{(2)} \notag\\
& + \left[D(\pt) -\abarmu{\pt} [D(\pt)]^{(1)}
  -\left(\abarmu{\pt}\right)^2 [D(\pt)]^{(2)}  \right]+ {\rm
  regular~terms~of~{\cal O}(\as^3)}\bigg\}.
\end{align}
The NNLO fully differential cross section is then obtained upon
integration over $\pt$ from scales of the order of the Landau pole
$\Lambda$ to the kinematic upper bound.  Each term in
Eq.~\eqref{eq:minnlo} contributes to the total cross section according
to the power counting formula
\begin{equation}
    \int_{\Lambda}^{Q} \mathd \pt \frac{1}{\pt} \as^m(\pt) \ln^n\frac{Q}{\pt}
\exp(-\tilde{S}(\pt))    \approx {\cal O}\left(\as^{m-\frac{n+1}{2}}(Q)\right)\,, 
\label{eq:counting}
\end{equation}
which clarifies why certain terms of ${\cal O}(\as^3)$ should be
included.

The above considerations can be made at the fully differential level
on the $\PhiBJ$ phase space,  which leads to the definition of the
${\bar B}(\PhiBJ)$ function as~\cite{Monni:2019whf}
\begin{align}
\label{eq:Bbar}
      {\bar B}(\PhiBJ)& \equiv \exp[-\tilde{S}(\pt)]\Bigg[ B(\PhiBJ)\left(1+\abarmu{\pt}\left[\tilde{S}(\pt)\right]^{(1)}\right) + V(\PhiBJ)\nonumber \\
       &  + D(\pt)^{\rm (\ge 3)} F^{\rm corr}(\PhiBJ)   \Bigg] + \int d\Phi_{\rm rad} R(\PhiBJ,\Phi_{\rm rad})\tilde{S}(\pt)\,, 
\end{align}
where $D(\pt)^{(\ge 3)}$ corresponds to the square bracket in the
second line of Eq.~\eqref{eq:minnlo}, and the factor
$F^{\tmop{corr}}(\PhiBJ)$ encodes its dependence upon the full
$\PhiBJ$ phase space, as discussed in detail in Section 3 of
\citere{Monni:2019whf}.

\subsection{Extension beyond the colour singlet case}
\label{sec:NNLOPStt}

In the case of the production of coloured final states, such as
top-quark pair production, the starting point to derive the singular term in
Eq.~\eqref{eq:start} is the more complex expression
\begin{equation}
\sum_{c=q,\bar{q},g}
  \frac{|M^{(0)}_{c\bar{c}}|^2}{2 m_{t\bar{t}}^2}\int\frac{d^2\vec{b}}{(2\pi)^2} e^{i \vec{b}\cdot
  \ptvec } e^{-S_c \left(\frac{b_0}{b}\right)}
\times\sum_{i,j}\Tr({\mathbf H}_c{\mathbf \Delta})\,
 \,({C}_{ci}\otimes f_i) \,({C}_{\bar{c} j}\otimes f_j) \,,
\label{eq:bspace}
\end{equation}
which describes the production of a pair of heavy quarks at small
transverse momentum. Here $b_0=2\,e^{-\gamma_E}$, $b=|\vec{b}|$. $S_c$
is the same Sudakov radiator which also enters the description of the
production of a colour singlet system at small transverse momentum.
The first sum in Eq.~\eqref{eq:bspace} runs over all possible flavour
configurations of the incoming partons $p_1$ of flavour $c$ and $p_2$
of flavour $\bar c$.
The collinear coefficient functions
$C_{ij}=C_{ij}(z,p_1,p_2,\vec{b};\alpha_s(b_0/b))$ describe the
structure of constant terms related to the emission of collinear
radiation, and the parton densities are denoted by $f_i$ and are
evaluated at $b_0/b$. The operation $\otimes$ denotes the standard
convolution over the momentum fraction $z$ carried by initial state
radiation.
The factor
$\Tr({\mathbf H}_c{\mathbf \Delta})\, \,({C}_{ci}\otimes f_i)
\,({C}_{\bar{c} j}\otimes f_j)$
has different expressions for the $q\bar q$ and $gg$ channels and has
here a symbolic meaning. In particular, it has a rich Lorentz
structure that we omit for simplicity, which
is a source of azimuthal correlations in the collinear
limit~\cite{Catani:2010pd,Catani:2014qha}.

All quantities in bold face denote operators in colour space, and the
trace $\Tr({\mathbf H}_c{\mathbf \Delta} )$ in Eq.~\eqref{eq:bspace}
runs over the colour indices.
The hard function ${\mathbf H}_c={\mathbf H}_c(\Phi_{\rm
  t\bar{t}};\alpha_s(m_{t\bar{t}}))$ is obtained from the subtracted
amplitudes and the ambiguity in its definition corresponds to using a
specific resummation scheme~\cite{Bozzi:2005wk}.
%We adopt here the definition of Ref.~\cite{Catani:2014qha}.
%
The operator ${\mathbf \Delta}$ encodes the structure of the quantum
interference due to the exchange of soft radiation at large angle
between the initial and final state, and within the final state. It is
given by ${\mathbf \Delta}={\mathbf V}^\dagger{\mathbf D}{\mathbf V}$,
where~\cite{Catani:2014qha}
\begin{align}
\label{eq:soft}
{\mathbf V} &= {\cal
  P}\exp\left\{-\int_{b_0^2/b^2}^{m_{t\bar{t}}^2}\frac{dq^2}{q^2}{\mathbf
  \Gamma}_t(\Phi_{\rm t\bar{t}};\alpha_s(q))\right\}\,.%\notag\\
\end{align}
The symbol ${\cal P}$ denotes the path ordering (with increasing
scales from left to right) of the exponential matrix with respect to
the integration variable $q^2$. ${\mathbf \Gamma}_t$ is the anomalous
dimension accounting for the effect of real soft radiation at large
angles, and 
${\mathbf D}={\mathbf D}(\Phi_{\rm t\bar{t}},\vec{b};\alpha_s(b_0/b))$
encodes the azimuthal dependence of the corresponding constant terms,
and is defined such that $[{\mathbf D}]_\phi={\mathbf{1}}$, where
$[\cdots]_\phi$ denotes the average over the azimuthal angle $\phi$ of
$\ptvec$.

The strategy to arrive at a \minnlo{} improved $\bar B$ function is
the same as for the colour-singlet case. We expand
Eq.~\eqref{eq:bspace} taking care of not spoiling the NNLO counting
accuracy outlined in Eq.~\eqref{eq:counting} and maintaining leading
logarithmic accuracy to arrive to an expression in transverse momentum
space that can be used to correct the $\bar B$ function in order to
achieve NNLO accuracy. All details are given in
Refs.~\cite{Mazzitelli:2020jio,Mazzitelli:2021mmm}.

\section{Selected results}
\label{sec:pheno}

Here, as an example, we discuss selected results for $ZZ$ and
$t \bar t$ production. The uncertainty bands shown are
in general obtained with a standard seven-point scale variation of
renormalisation and factorisation scales, and possibly with a
variation of the resummation scale, as described in the relevant
publications.

\subsection{$ZZ$ production}

The accurate simulation of the production of a pair of $Z$ bosons,
with subsequent decay into a four-lepton final state, is instrumental
for the precision program of the LHC. This process constitutes a relevant background in
Higgs boson measurements, and plays an important role in constraining
the presence of anomalous interactions in the gauge sector of the
Standard Model. Accordingly, an accurate event generation is of
paramount importance. In this section we present some sample \minnlo{}
results for this process.

\begin{figure*}[t!]
\includegraphics[width=0.49\textwidth]{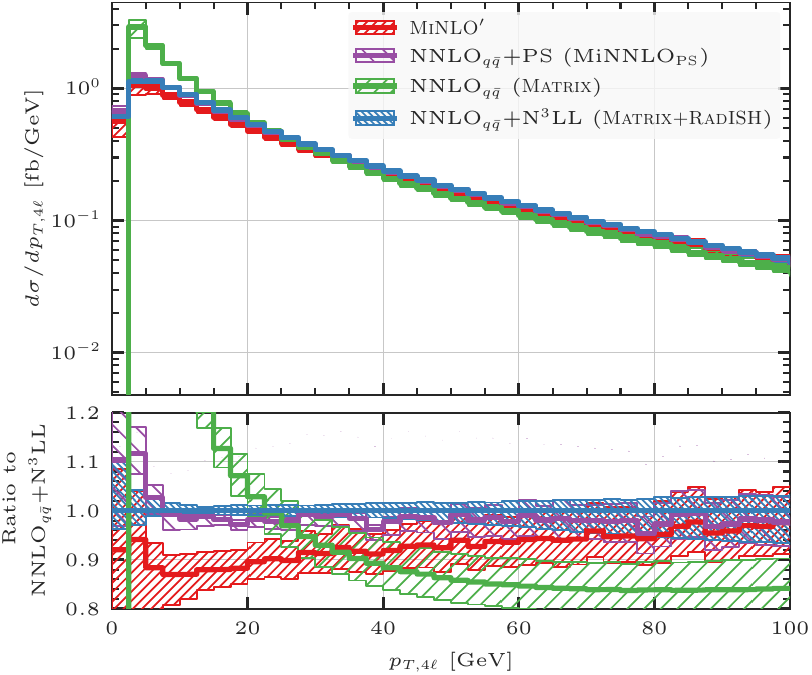}
\hspace{0.15cm}
\includegraphics[width=0.49\textwidth]{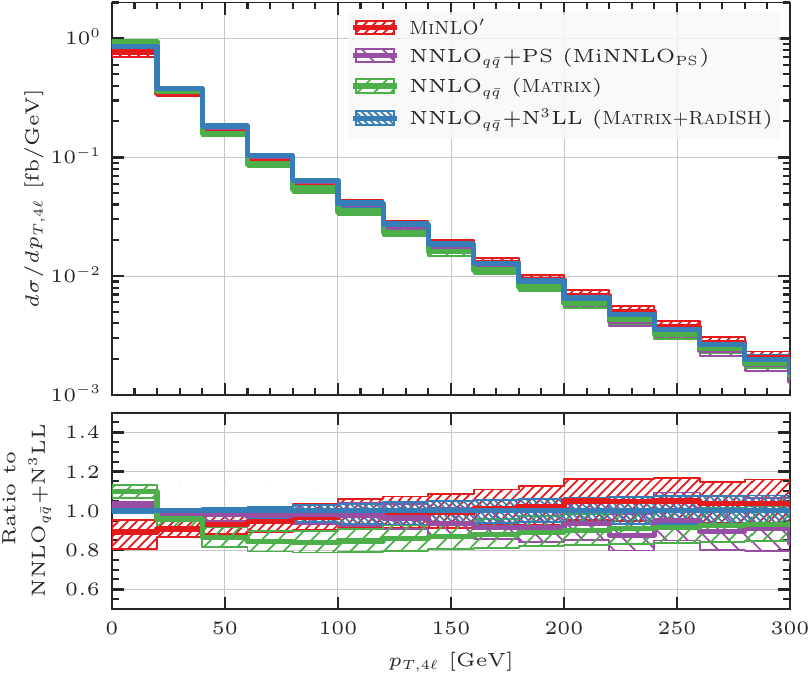}%
  \caption{Comparison between NNLO (\Matrix{}), \minlo, \minnlo and
    the NNLO+N$^3$LL of \Matrix{}+\radish~\cite{Kallweit:2020gva} for
    the transverse momentum of the $ZZ$ pair for two different ranges
    of $p_{T,4\ell}$. Figure taken from~\citere{Buonocore:2021fnj}. }
\label{fig:ZZ-pt4l}
\end{figure*}

\fig{fig:ZZ-pt4l}
displays predictions for the transverse momentum
of the diboson pair ($p_{T,4\ell}$).
We show \minnlo{} predictions (purple), compared to \minlo{} results (red). The latter is only NLO accurate,
and fixed-order NNLO predictions from \Matrix{}~\cite{Grazzini:2017mhc} (green). 
We also show the NNLO+N$^3$LL result obtained with
\Matrix{}+\radish~\cite{Kallweit:2020gva} (blue), which interfaces
\Matrix{} to the \radish{} resummation formalism
\cite{Monni:2016ktx,Bizon:2017rah}.
Since \Matrix{}+\radish{} does not include the contribution stemming from the
loop-induced $gg$ channel, we perform this comparison by considering
in our \minnlo{} result only the $q \bar q$-initiated process, i.e.\ at the
NNLO$_{q\bar{q}}$+PS level.  At small values of the $ZZ$ transverse
momentum we observe an excellent agreement between the NNLO+N$^3$LL
and the \minnlo{} result, especially considering the lower accuracy of
the parton shower in that region; \minnlo{} is between $5$\% and $12$\%
larger than the NNLO+N$^3$LL prediction below 10 GeV and has a larger
uncertainty band reflecting its lower accuracy.  On the other hand,
the \minlo{} result is $\mathcal O(10 \%)$ smaller than the NNLO+N$^3$LL
and the \minnlo{} predictions, and its uncertainty band does not overlap with
either of the more accurate results below 40\,GeV.  Fixed-order
calculations actually lead to unphysical results in the
small-$p_{T,4\ell}$ region due to large logarithmic corrections, which
need to be resummed to all orders.  Indeed, the NNLO result diverges
at low transverse momentum, and its prediction differs significantly
from the ones including resummation effects.  At larger values
of $p_{T,4\ell}$ the NNLO result is instead in agreement with the
NNLO+N$^3$LL, \minlo{} and \minnlo{} predictions, as one may expect since
all of them have the same formal accuracy in the tail of the
distribution.

\begin{figure*}[tbh]
\includegraphics[width=0.48\textwidth]{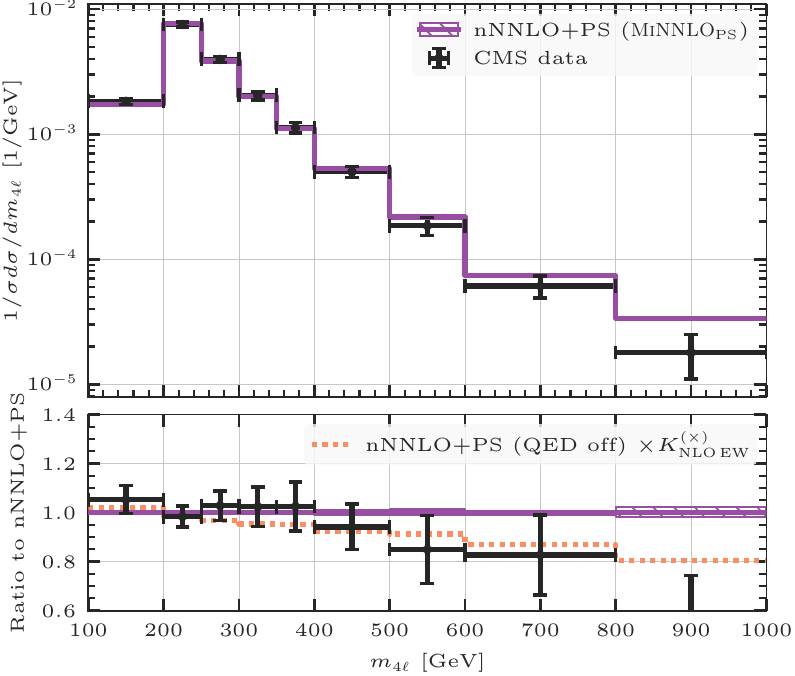}%
\hspace{0.3cm}
\includegraphics[width=0.48\textwidth]{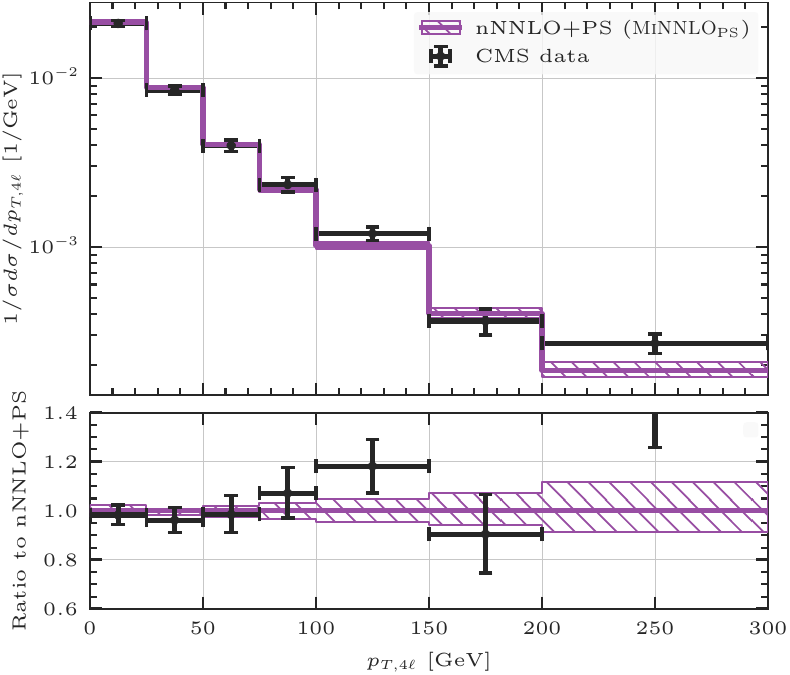}%
\vspace{0.2cm}
  \caption{Comparison between the \minnlo{} predictions and the CMS data
    of \citere{CMS:2020gtj} based on a 137 fb$^{-1}$ 13\,TeV analysis
    for $m_{4\ell}$ and $p_{T,4\ell}$. The \minnlo{} predictions include
    hadronization and multi-parton interactions effects, as well as QED effects as provided
    by the \PYTHIA{8} parton shower. Figure taken from~\citere{Buonocore:2021fnj}.}
	\label{fig:data-minnlo}
\end{figure*}

Next we show a comparison to CMS data of~\citere{CMS:2020gtj} for the
invariant mass and the transverse momentum of the diboson pair
($m_{4\ell}$ and $p_{T,4\ell}$), now including also the gluon-induced
production described at NLOPS accuracy.
By and large, we observe a quite good agreement between our
predictions and the experimental data.  The invariant mass is well
described at low $m_{4\ell}$, but there is a tendency of the
predictions to overshoot the data at large $m_{4\ell}$, with the last
bin being almost two standard deviations away. In this region
electroweak (EW) corrections are known to be important and they are
only partly included here through the QED shower. A simple inclusion
of the NLO EW corrections, obtained via a reweighting of the events by a proper
EW K-factor (see~\citere{Buonocore:2021fnj} for more details), can
visibly improve the agreement with data in this region. The
transverse-momentum distribution of the $ZZ$ pair is also well
described, except for a two-sigma deviation in the last bin, with a
remarkable agreement for $p_{T,4\ell}$ values below $ \sim 100$\,GeV,
where the all-order corrections provided by the shower are
particularly important.

\subsection{$t \bar t$ production}

The production of a pair of top quarks enters a number of important
analyses at the LHC. Due to its abundance, it is an important
background to several electroweak and Higgs processes, as well as
beyond the Standard Model searches. Moreover, it constitutes the main
production mechanism of top quarks at the LHC, and it is therefore an
ideal process to study top-quark properties accurately.

\begin{figure}[t!]
\begin{center}
\begin{tabular}{ccc}
\hspace{-.55cm}
\includegraphics[width=.43\textwidth,page=1]{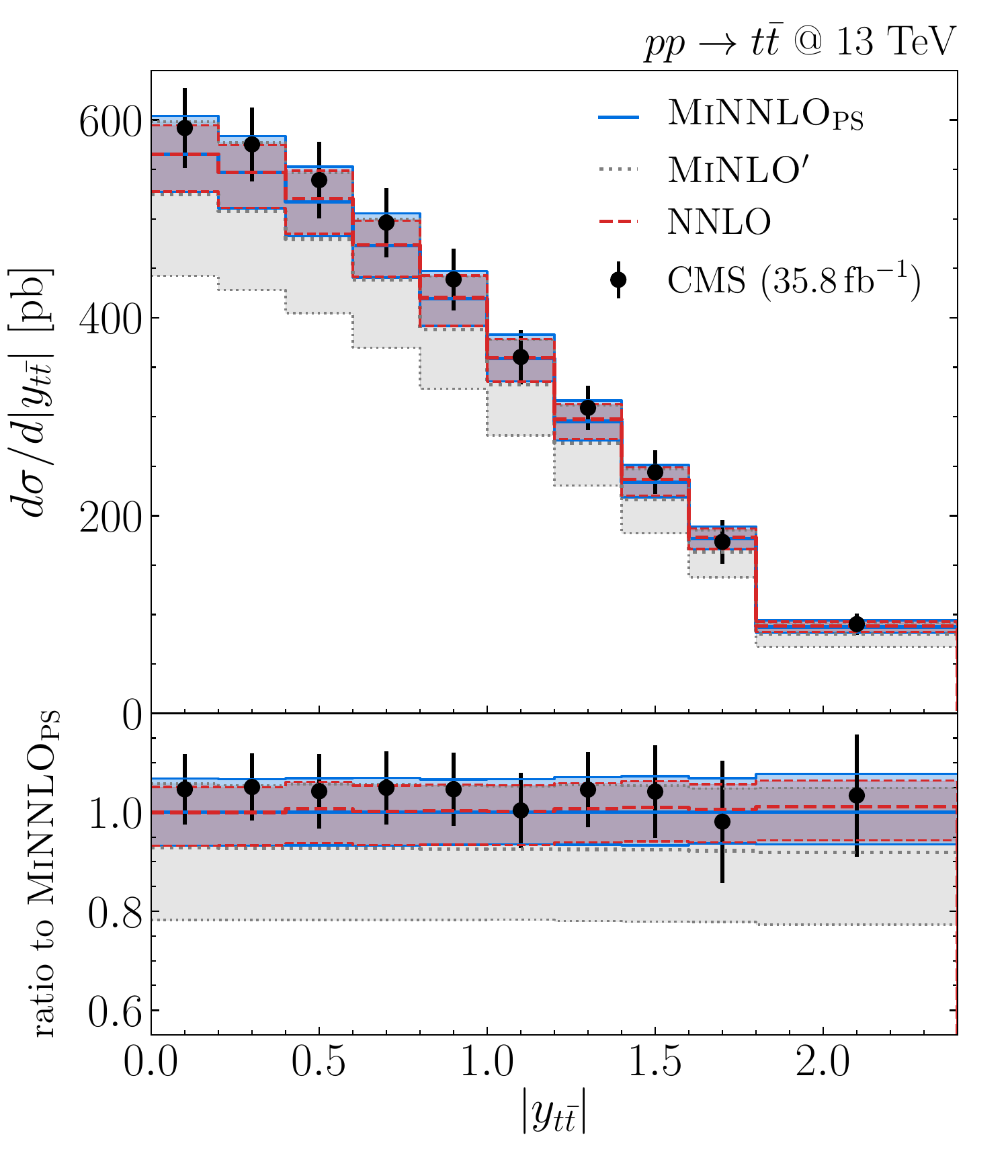}
&
\includegraphics[width=.43\textwidth,page=6]{all_plots_paper.pdf} 
\end{tabular}
\caption{\label{fig:tt-inclusive} Comparison of \minnlo{} (blue, solid),
  \minlo{} (black, dashed), and NNLO QCD (red, dashed) predictions
  with CMS data \cite{CMS:2018htd} (black points with errors) with no
  phase space cuts. Figure taken from Ref.~\cite{Mazzitelli:2021mmm}.}
\end{center}
\end{figure}

In \fig{fig:tt-inclusive}, 
we show \minnlo{} (blue, solid), \minlo{} (gray, dotted) and
fixed-order NNLO (red, dashed) predictions compared to data from CMS
\cite{CMS:2018htd} (black points with errors) that has been
extrapolated from semi-leptonic top-quark decays to the inclusive
$t\bar{t}$ phase space. We show the rapidity $y_{t\bar{t}}$ and the
transverse momentum of the leading top $p_{T,t_1}$.
With respect to \minlo{}, the \minnlo{} corrections lead to a
significant increase of about $10$\% in the central value, and a
substantial reduction of the scale uncertainties of more than a factor
of two.
We find excellent agreement between \minnlo{} and fixed-order NNLO
predictions for the rapidity distribution, both for the central
predictions and for the scale-uncertainty bands.
For the transverse momentum distribution we observe mild differences
between \minnlo{} and NNLO at small $p_{T,t_1}$, especially in terms
of shape. This is expected since this distribution is affected by
large logarithmic contributions in this region, therefore a matching
to the parton shower becomes particularly important. The observed
differences with respect to NNLO are however largely covered by the
perturbative uncertainties and both results are in excellent agreement
with the experimental data.
%

%=================================================================
We continue by considering \minnlo{} predictions in comparison to data
for various distributions in the phase space of the top-decay
products, now including fiducial selection cuts.  For simplicity, we
only show the \minlo{} results as a reference prediction in the
following.
We start by considering the decay channel where both top quarks decay
leptonically, requiring one electron and one muon in the final
signature.
\begin{figure}[t!]
\begin{center}
\begin{tabular}{cccccc}
\hspace{-0.5cm}
\includegraphics[height=.32\textheight,page=1]{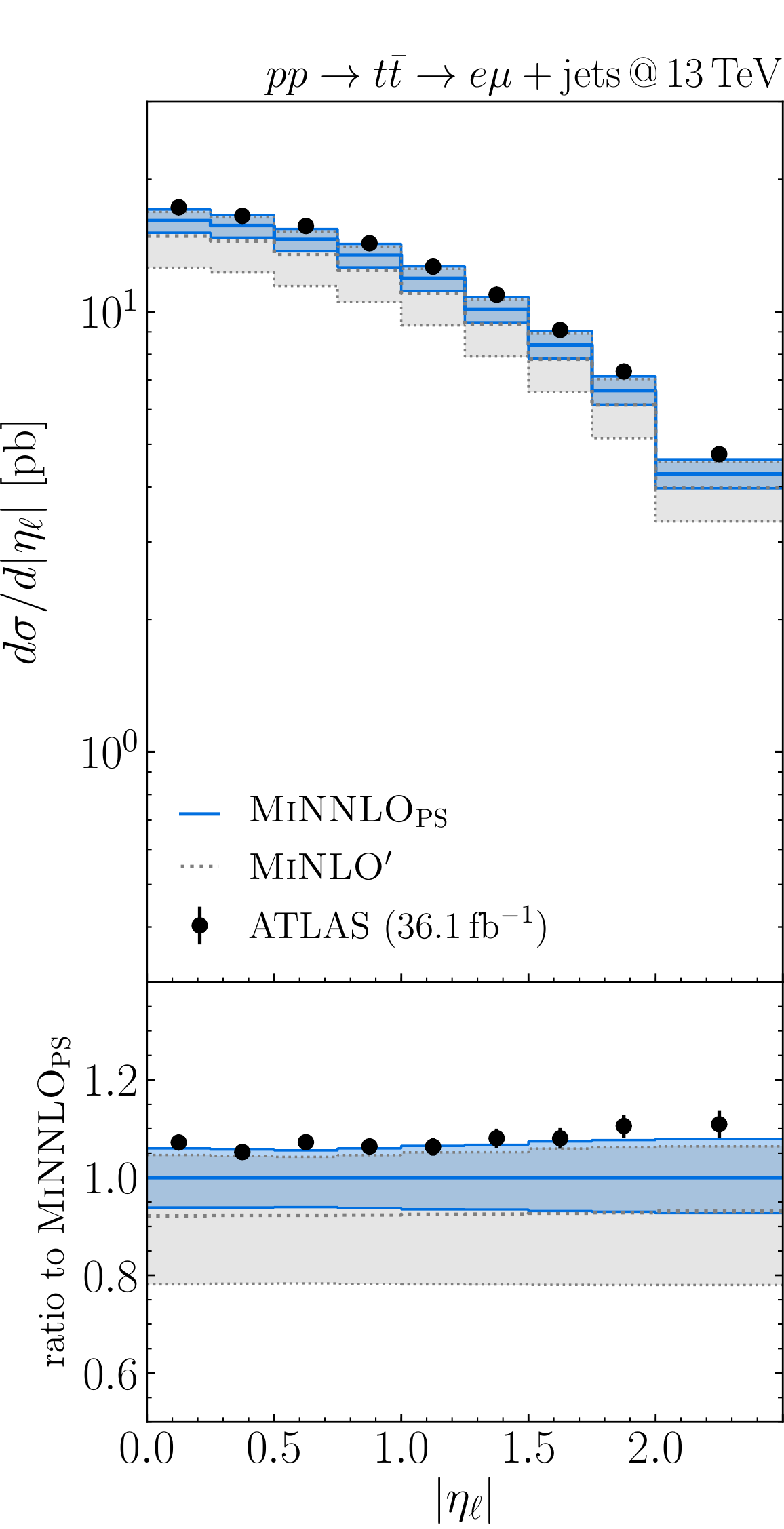}
&
\hspace{-0.5cm}
\includegraphics[height=.32\textheight,page=2]{ATLAS_leptonic_with_tau.pdf}
&
\hspace{-0.5cm}
\includegraphics[height=.32\textheight,page=3]{ATLAS_leptonic_with_tau.pdf} 
&
\hspace{-0.5cm}
\includegraphics[height=.32\textheight,page=4]{ATLAS_leptonic_with_tau.pdf}
&
\hspace{-0.5cm}
\includegraphics[height=.32\textheight,page=5]{ATLAS_leptonic_with_tau.pdf}
\end{tabular}
\caption{\label{fig:leptonic} Comparison of \minnlo{} (blue, solid)
  and \minlo{} (gray, dashed) predictions with ATLAS data
  \cite{ATLAS:2019hau} (black points with errors) in the fully
  leptonic decay mode, including decays of $\tau$ leptons.}
\end{center}
\end{figure}
In Figure~\ref{fig:leptonic} we show the rapidity of the two leptons
$\eta_{\ell}$ (binned together), inclusively over $m_{e\mu}$ (first
panel) as well as in slices of $m_{e\mu}$, as measured by ATLAS at
$13$\,TeV~\cite{ATLAS:2019hau} (black points with errors) in
comparison to \minnlo{} (blue, solid) and \minlo{} (gray, dotted)
predictions.
In the results considered here, the electrons and muons may also stem
from top-quark decays to $\tau$ leptons and their subsequent leptonic
decays.
The shapes are described very well and the data is at the upper edge
of the uncertainty band of \minnlo{}, but overall still
compatible. Compared to the \minlo{} prediction, we observe a
substantial reduction of the uncertainty band.

\section{Outlook}

The \minnlo{} method can currently be used to describe processes with
generic colour structure that, at the Born level, do not include light
partons in the final state, the only requirement being the availability
of the resummation of a suitable kinematic variable (such as, for
instance, the transverse momentum of the heavy system) at the desired
perturbative accuracy as well as the NNLO computation of the hard
process.
The next milestone is likely to be the matching for a process with a
light parton in the final state, such as Higgs plus jet or Drell-Yan
plus jet. Similarly, NLO EW corrections may compete with NNLO QCD ones
and a joint inclusion of both is another important future goal.
A crucial observation is that the ongoing efforts to improve the
logarithmic accuracy of the shower to NLL, and eventually NNLL level,
are likely to add new requirements for NLOPS and NNLOPS
methods, which at the moment only preserve the LL accuracy of the
shower.

\bibliographystyle{apsrev4-1}
\bibliography{MiNNLO}
\end{document}